# Integrated Impact Indicators (*I3*) compared with Impact Factors (*IFs*):

# An alternative research design with policy implications

Loet Leydesdorff [i] and Lutz Bornmann [ii]

**Abstract**

In bibliometrics, the association of "impact" with central-tendency statistics is mistaken. Impacts add up, and citation curves should therefore be integrated instead of averaged. For example, the journals *MIS Quarterly* and *JASIST* differ by a factor of two in terms of their respective impact factors (*IF*), but the journal with the lower *IF* has the higher impact. Using percentile ranks (e.g., top-1%, top-10%, etc.), an integrated impact indicator (*I3*) can be based on integration of the citation curves, but after normalization of the citation curves to the same scale. The results across document sets can be compared as percentages of the total impact of a reference set. Total number of citations, however, should not be used instead because the shape of the citation curves is then not appreciated. *I3* can be applied to any document set and any citation window. The results of the integration (summation) are fully decomposable in terms of journals or institutional units such as nations, universities, etc., because percentile ranks are determined at the paper level. In this study, we first compare *I3* with *IFs* for the journals in two ISI Subject Categories ("Information Science & Library Science" and "Multidisciplinary Sciences"). The LIS set is additionally decomposed in terms of nations. Policy implications of this possible paradigm shift in citation impact analysis are specified.

[i] University of Amsterdam, Amsterdam School of Communication Research (ASCoR), Kloveniersburgwal 48, 1012 CX Amsterdam, The Netherlands; loet@leydesdorff.net.
[ii] Max Planck Society, Hofgartenstrasse 8, D-80539 Munich, Germany; bornmann@gv.mpg.de.

**Introduction to the problem**

Let us introduce the problem of defining impact by taking as an example the citation curves of two journals with very different impact factors (*IF*s):

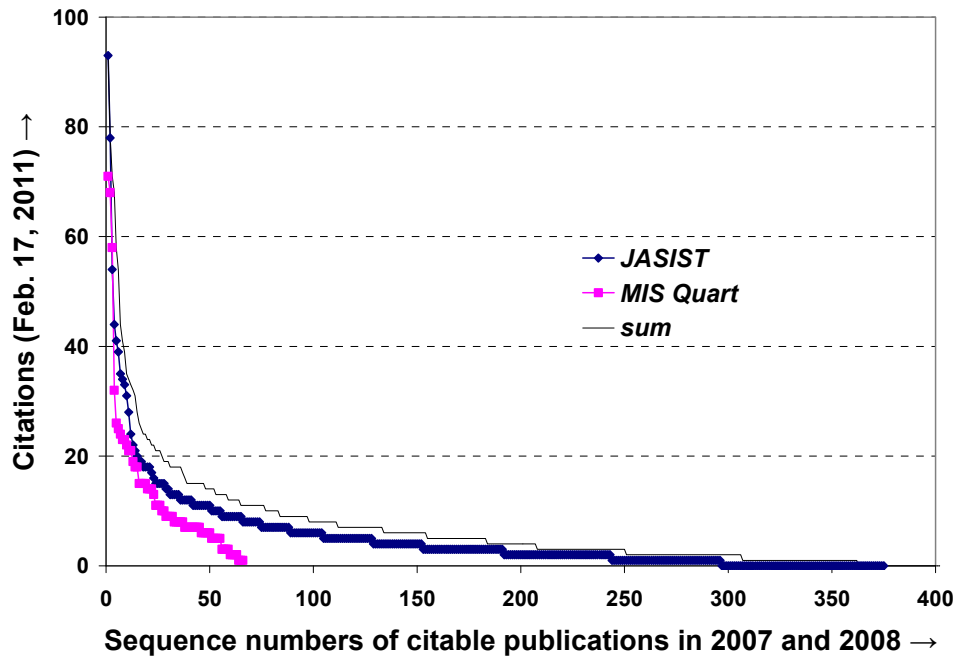

**Figure 1**: Citation curves for *JASIST* (n = 375 publications) and *MIS Quarterly* (n = 66).

Figure 1 shows the citation curves of the 66 and 375 citable items published in *MIS Quarterly* and *JASIST*, respectively, during 2007 and 2008.[1] These two journals are both attributed by Thomson Reuters—the present owner of the Institute of Scientific Information (ISI)—to the ISI Subject Category of "Library and Information Science" (LIS),[2] although they are very different in character (Nisonger & Davis, 2005; Zhao & Strotman, 2008). Within this Subject Category, *MIS Quarterly* had the highest *IF* in 2009: 4.485. The IF of *JASIST* is approximately half this size: IF-2009 = 2.300. However, the 66 most-highly cited publications of *JASIST* obtained 380 citations more than the 66 citable items published in *MIS Quarterly* (downloaded on February 17, 2011). The lower *IF* factor is entirely due to the tail of 300+ additional publications in *JASIST* with lower citation rates.

---

[1] The *Journal Citation Reports* (JCR) 2009 lists 370 instead of 375 citable issues for 2007 plus 2008. This difference originates from the date in March that the JCR-team at Thomson Reuters decides to use for producing the JCR of the year before (McVeigh, *personal communication*, April 7, 2010). *IF*s are notoriously difficult to reproduce using Web-of-Science data (e.g., Brumback, 2008a and b; Rossner *et al.*, 2007 and 2008; Pringle, 2008).

In our opinion, this confusion finds its origin in the definition of the "impact factor" as a two-year average of "impact" (Garfield, 1972; Garfield & Sher, 1963; cf. Bensman, 2008; Rousseau & Leydesdorff, 2011).[3] Impact (as a variable), however, is not an average, but the result of the sum of the momenta of the impacting units. For example, two meteors impacting on a planet can have a combined impact larger than that of each of them taken separately, but the respective velocities also matter.

In physics, momentum is defined as the vector of mass times velocity ($p = m\vec{v}$). Using the metaphor of impact, both the number of publications (the "mass") and their citation counts (the quality of "the velocity") matter for the impacts. Because citations are scalar counts, one can disregard the direction of the vectors in the summation ($\Sigma\, m\vec{v}$). The research question is then how to operationalize *m* in terms of the numbers of publications and *v* in terms of citations in order to obtain a relevant measure of impact as a sum. The impact of each subset can be expressed as a percentage impact of the set.

---

[2] The ISI uses "Information Science & Library Science" as name of this category.
[3] More recently, the ISI also introduced the five-year IF in the Journal Citation Reports (JCR).

**Table 1**: Comparison of *MIS Quarterly* and *JASIST* in terms of citation rates to citable items in 2007 and 2008. Publication and citation data retrieved at the Web of Science (WoS) on February 17, 2011.

|  | IF 2009 | (P)ublications in our data | (C)itations in our data | C/P | Median |
|---|---|---|---|---|---|
| *MIS Quart* | 296/66 = 4.485 | 66 | 847 | 12.83 | 8 |
| *JASIST* | 851/370 = 2.300 | 375 | 1975 | 5.27 | 3 |
| sum | 1147/436 = 2.631 | 441 | 2822 | 6.40 | 3 |

It has been argued (e.g., Bornmann & Mutz, 2011; Leydesdorff & Opthof, 2011) that the median should be used in citation analysis instead of the mean because of the skewness of citation distributions (e.g., Seglen, 1992). For the two journals in the example above—and using the two-year time window of the ISI-*IF*—Table 1 shows that the median is even more sensitive to the tails of the distributions than the mean. A more radical solution is therefore needed: impact has to be defined not as a distribution, but as a sum. Very different distributions can add up to the same impact. The number of citations can be highly skewed and in this situation any measure of central tendency is theoretically meaningless.

Whereas distributions of citations can be tested non-parametrically for the significance of the differences among them, impacts are sum values. These values can be tested against the expected values of the variables. For example, if the set of documents in one journal is twice as large as the set in another, the chance that it will contain a top-1% most highly-cited document is twice as high. If the observed value, however, would be four times as high, this achievement above expectation may be statistically significant, but this depends also on the sample size ($N$).

Central tendency statistics cannot capture the increases in impact when two sets ("masses") are added; for example, when two research groups join forces or two journals merge. We penciled the line for the sumtotal of *JASIST* and *MIS Quarterly* into Figure 1 in order to show that one has to sum surfaces, and thus the citation curves have to be *integrated* instead of averaged. Assuming the integrals, however, would lead to numbers equal to the "total citations" without qualifying the documents in terms of their citedness. In order to weigh the documents, we suggest to transform the citation curves first into curves of hundred percentiles as in Figure 2.

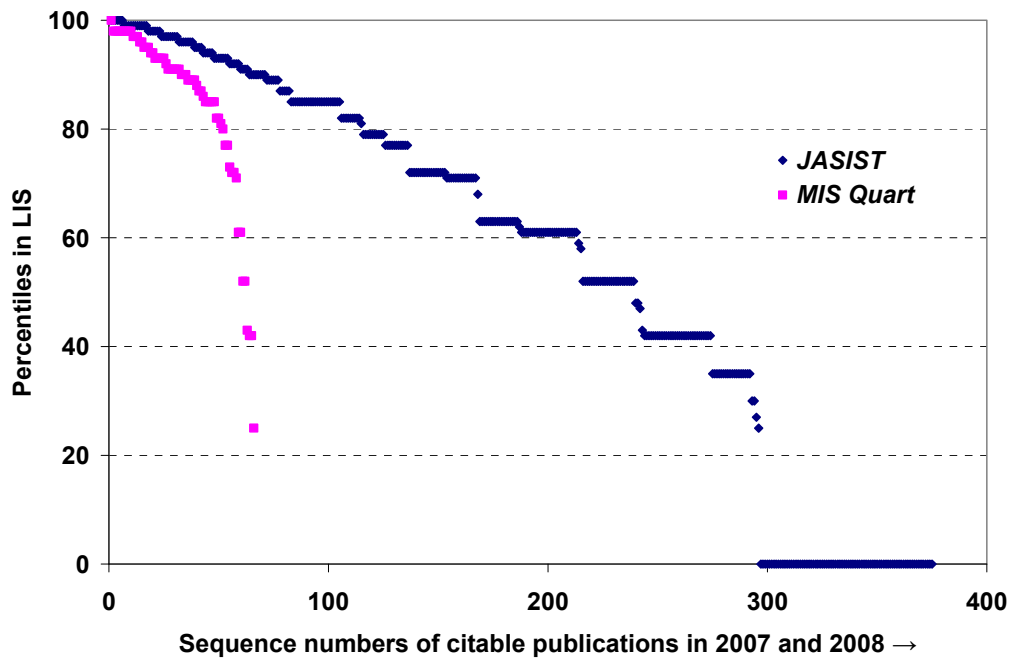

**Figure 2**: Distributions of 100 Percentile Ranks of *JASIST* and *MIS Quarterly* with reference to the 65 journals of the ISI Subject Category LIS.

The distributions of percentile ranks can fairly be compared across document sets, and these linearly transformed distributions can be integrated. The integrals in this stepwise function are equal to $\sum_i x_i * f(x_i)$ in which *x* represents the percentile rank and *f*(x) the frequency of that rank; *i* is one hundred when using percentiles, but, for example, four when using quartiles as percentile rank classes (etc.). One can also consider hundred percentiles as a continuous random variable and sum these values, as we shall explain in more detail in the methods section below. The function integrates both the number of papers (the "mass") and their respective quality in terms of being-cited normalized as percentiles with reference to a set.

The idea of using percentile rank classes was first formulated in the discussion about proper normalization of the citation distribution that took place last year in the *Journal of Informetrics* (e.g., Bornmann, 2010; Opthof & Leydesdorff, 2010; Van Raan *et al*., 2010a; cf. Gingras & Larivière, 2011). In this context, one of us proposed to assess citation distributions in terms of six percentile rank classes (*6PR*): the top-1%, top-5%, top-10%, top-25%, top-50%, and bottom-50% (Bornmann & Mutz, 2011). This (normative!) evaluation scheme accords with those currently used in the bi-annual *Science & Technology Indicators* of the National Science Board of the USA (2010, at Appendix Table 5-43). Each publication would then be weighted in accordance to its class as a six for the top-1% category and a one for the bottom-50% category. Leydesdorff, Bornmann, Mutz & Opthof (in press) extended this approach to hundred percentiles which can also be weighted as classes from 1 to 100 (*100PR*).

The advantage of using percentile ranks is that one is thus able to compare distributions of citations across unequally sized document sets using a single scheme for the evaluation of the shape of the distribution. However, Bornmann and Mutz's (2011) approach remained sensitive to the central-tendency characteristic discussed above because these authors *averaged* over the percentile ranks using the following formula: $R_i = \sum_i x_i * p(x_i)$. In this formula, *x* is the rank class and *p(x)* its *relative* frequency (or proportion). However, this probabilistic approach implies a division by the *N* of cases and thus normalization to the mean (albeit of the distribution of the percentiles). Leydesdorff *et al.* (in press) therefore called this method the "mean percentile rank" approach.

Means cannot be added, whereas impacts are additive. Bensman & Wilder (1998) have concluded on the basis of validation studies that the prestige of journals in chemistry is correlated with the total number of citations more than with the impact factors of journals. As noted above, however, total citations do not yet qualify the shape of the distributions since every citation is then counted equally. Impact factors only qualify the distribution in terms of the mean, but deliberately abstract from size (Bensman, 2008). Using the sum total of the frequencies (*f*) in each percentile, however, accounts for both the size and shape of the distribution. The citations are weighted in accordance with the percentile rank class of each publication in the Integrated Impact Indicator: $I3 = \sum_i x_i * f(x_i)$.

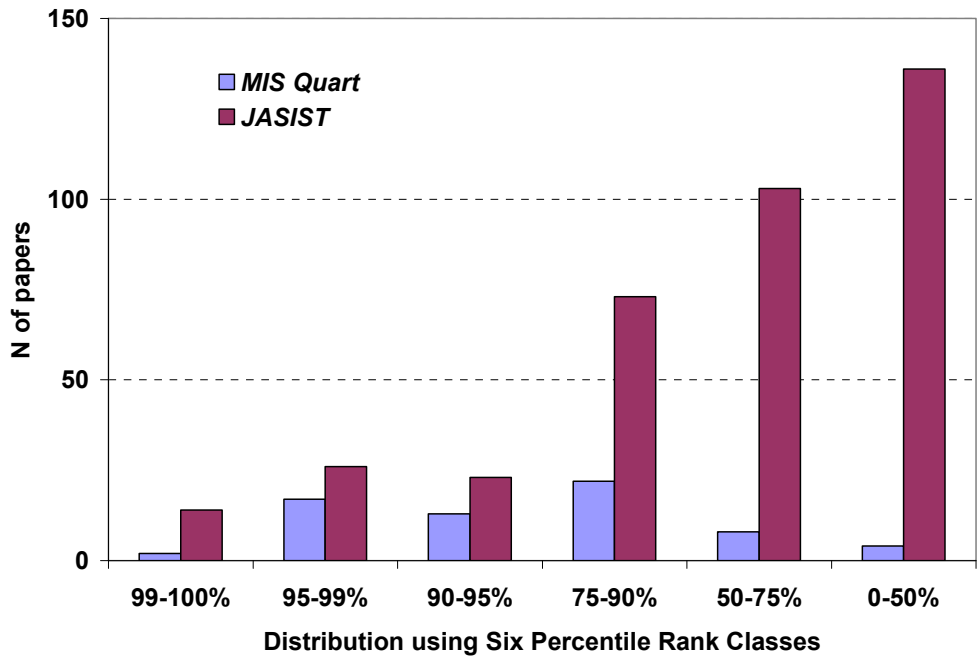

**Figure 3**: Distributions of the six percentile ranks of publications in terms of citations to *JASIST* and *MIS Quarterly* (with reference to all 65 of LIS).

Figure 3 shows the distribution of the six percentile ranks of the National Science Board (2010) for *MIS Quarterly* and *JASIST*. Both Figures 2 and 3 show that *JASIST* has an impact higher than *MIS Quarterly* using these normalized curves. Table 2 shows that this higher value can be captured by the sum, but not by the means or medians of the percentile distributions.

**Table 2**: Mean (± s.e.m.), median and sum in the case of hundred or six percentile ranks.

|  | 100PR | | | 6PR | | |
| --- | --- | --- | --- | --- | --- | --- |
|  | Mean | Median | Sum | Mean | Median | Sum |
| *MIS Q* | 84.57 ± 1.98 | 90 | 5,581.4 | 3.56 ± 0.15 | 3 | 235 |
| *JASIST* | 55.50 ± 1.76 | 61 | 20,811.3 | 2.31 ± 0.07 | 2 | 867 |

The sum values can be added (and subtracted), for example, for the purpose of the aggregation or decomposition (e.g., in terms of contributing nations), and they can also be expressed as percentages of the total integrated impact of an ISI Subject Category (e.g., the 65 journals in the LIS category). For example, the sum total of the impact of all 5,737 citable items in the LIS category was in 2009: 213,906.2.[4] *MIS Quarterly* contributed 2.61% to this total impact of the set when using *100PR*, and 2.34% in the case of *6PR*. These percentages were 9.73% and 8.63% for *JASIST*, respectively. *JASIST* is thus to be considered as the journal with the highest impact in the set of 65 journals subsumed under LIS among the ISI Subject Categories (Nisonger, 1999).

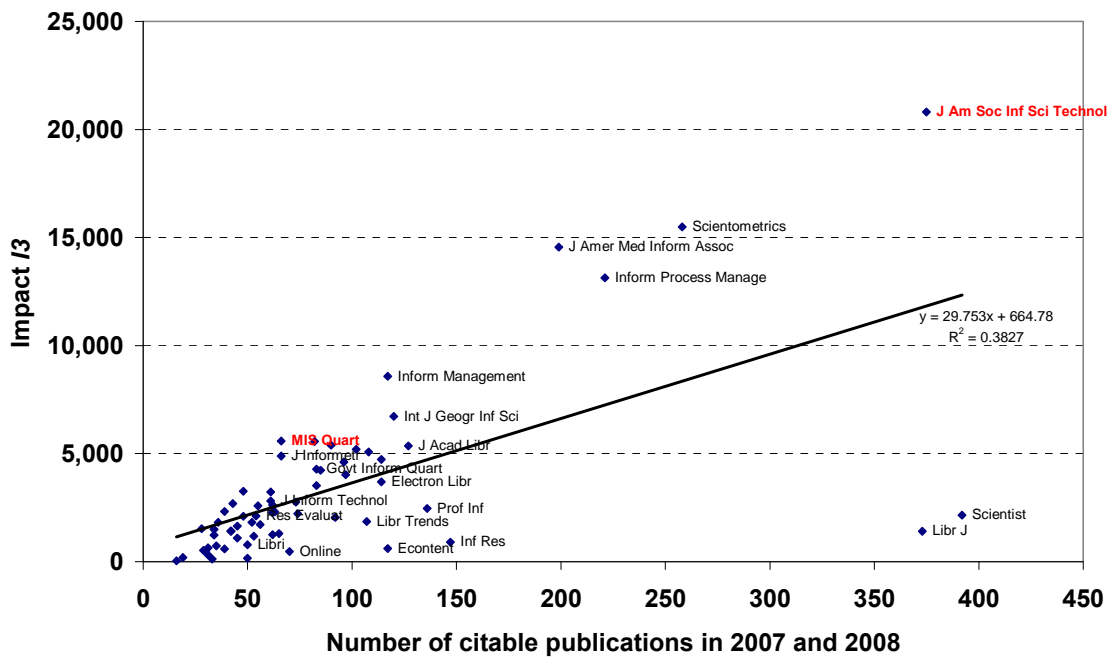

**Figure 4**: Regression of impact (*I3*) against number of citable publications in 2007 and 2008 for the 65 journals of LIS.

Figure 4 further elaborates this example by showing the regression line of the impacts thus calculated against the number of citable publications in 2007 and 2008. Unlike dividing sums to obtain average values—which was the core issue in the previous controversy about the Leiden crown indicator (Gingras & Larivière, 2011)—this regression informs us that the journal set under study is heterogeneous ($r^2 = 0.38$), both in size and functions. The *Scientist* and the *Library Journal*, for example, are grouped in the bottom-right angle of this figure because they function as newsletters more than scholarly journals. Among the journals at the top right, the label for *JASIST* was colored red in order to show its leading position in this set. *MIS Quarterly* is a journal above the regression line in the set of specialized journals at the bottom left, but also colored red for the purpose of this comparison.

---

[4] This sum total is equal to 39.8% of the maximally possible impact of (100 * 5,737 =) 537,700 in the case of *100PR*. In the case of *6PR*, the total is 10,049 or 29.2% of the maximally possible impact of (6 * 5,737 =) 34,422.

**Methods**

Data was harvested from the WoS in February 2011. Because we wished to compare our results for *I3* with the latest available *IFs* 2009, we downloaded citable items in 2007 and 2008 in two ISI Subject Categories, namely, the one for LIS containing 65 journals and the category "Multidisciplinary Sciences" (MS) containing 48 journals, but including important journals such as *Science*, *Nature*, and *PNAS*. The delineation of the ISI Subject Categories is beset with error (Rafols & Leydesdorff, 2009). Within this context, however, we use them pragmatically as reference sets because it is beyond our capacity—and perhaps illegal—to download the entire database. Only articles, proceedings papers, reviews, and letters are included because these categories are indicated by Thomson Reuters—the present owner of the Institute of Scientific Information (ISI)—as citable items (cf. Moed & Van Leeuwen, 1996). Note that *I3* is by no means confied to this definition of impact in terms of two previous years, but can be used for any document set and with any citation window.

The citation of each paper is rated in terms of its percentile in the distribution of citations to all items with the same document type and publication year in its ISI Subject Category as the respective reference sets. The percentile is determined by using the counting rule that the number of items with lower citation rates than the item under study determines the percentile. Tied citation numbers are thus provided with the highest values, and this accords with the idea of providing all papers with the highest possible ranking (in other words: we wish to give the papers "the benefit of the doubt"). Other schemes are also possible. For example, Pudovkin & Garfield (2009) first averaged tied ranks.

The percentiles can be considered as a continuous random variable in the case of one hundred percentiles. In the case of six percentile ranks, one has to round off. Differently from Leydesdorff *et al*. (in press), the rounding off will in this study be based on adding 0.9 to the count— that is: (count + 0.9)—because otherwise one can expect undesirable effects for datasets that are smaller than one hundred. For example, if a journal with many articles publishes only 10 reviews each year, the highest possible percentile within this set would be the 90$^{th}$—nine out of ten—whereas this could be the 99$^{th}$—that is, 9.9 out of 10, and thus included in the top-1% percentile rank (with a value of six in the mentioned evaluation scheme of the NSF).

As shown in Figures 2 and 3 above comparing *MIS Quarterly* with *JASIST*, the percentiles provide us with a scale that can be compared across document sets with different sizes. In the case of a normative evaluation scheme such as that of the NSF, the percentiles are binned in six percentile rank classes. This transformation is non-linear and one looses information, but an evaluator may gain clarity in the distinctions from a policy perspective (Leydesdorff *et al.*, in press). We shall use this second set of values throughout this study for the comparison, but distinguish it from *I3* by denoting this measure as *I3*(*6PR*). The formula is then specified as follows: $I3(6PR) = \sum_{i=1}^{6} x_i * PR_i$ , in which $PR_i$ is the frequency value in the respective class. Other evaluation schemes are also possible, but this is, in our opinion, a normative discussion which can be expected to change with the policy context.

The set based on the 65 journals of LIS contained 5,737 citable items published in 2007 or 2008. The set indicated by the ISI as journals in MS was much larger in terms of the number of documents (24,494 citable items) despite the smaller number of journals (48). The two sets were brought under the control of relational database management and when necessary dedicated routines were written in order to format the data for analysis in SPSS (v. 18) and Excel. Using the WoS, the numbers of citations were determined at the date of downloading, in our case February 2011.

The most relevant routine in SPSS is "Compare Means" using the journals (in each set, respectively) as the independent (grouping) variable and the percentiles (or the six classes, *mutatis mutandis*) as the dependent variables. This routine allows for determining the mean, the sum, the standard error of the mean, confidence levels, and other statistics in a single pass. Since we are mainly interested in the sum, the mean, and the standard error of the mean, this routine is sufficient for our purpose. Correlation analysis (both Pearson's $r$ and Spearman's rank-order correlation $\rho$) will also be pursued using SPSS in order to compare the new indicators with *IF*s.

In addition to analyzing the impact of each journal, the question can be raised of whether the citation distributions are also significantly different. Non-parametric statistics enable us to answer this question using the citation distributions (as depicted in Figure 1) without averaging or first transforming them into percentile ranks. Among the routines available for multiple comparisons in SPSS (with Bonferroni correction), Dunn's test can be simulated by using LSD ("least significant differences") with family-wise correction for Type-I error probability. In the case of N groups to be compared, the number of comparisons is $N * (N-1) / 2$. For example, in the case of 50 journals, $50 * 49 / 2 = 1,225$ comparisons are pursued, and the significance should hence be tested at the five percent level using $0.05 / 1,225 = 0.000041$ instead of 0.05 (Levine, 1991, at pp. 68 ff.).

The routine for multiple comparisons in SPSS is limited to 50 groupings at a time. In the case of the MS set, 48 journals are involved, but in the case of LIS 65 journals can be compared. We perform the analysis in this study using the 50 journals with the highest *IF*s among these 65 journals. (The *IF* was chosen as criterion in order not to bias our results in favour of the proposed measure.) Alternatively, one can test any two journals against each other using the Mann-Whitney U test with Bonferroni correction. However, in the case of 65 journals, these would be 2,080 one-by-one comparisons. This seemed not necessary for the purpose of this study.

Furthermore, the algorithm of Kamada & Kawai (1989)—as available, for example, in Pajek—provides us with a means to visualize groups of journals as *not* significantly different—or, in other words, homogenous—in terms of their citation distributions (cf. Leydesdorff & Bornmann, 2011, at p. 224f.). Journals which can be considered similar in this respect were linked in the graphs, while in the case of significant differences the grouping links were omitted. The *k*-core sets which are most homogeneous in terms of citation distributions can thus be visualized.

In a final section, we return to the issue of performance measurement of institutional units, individuals, and/or nations (but using this same data). Since the attribution of the percentile rank is done at the paper level, one can aggregate and decompose sub-sets in terms of their contribution to the reference set. We shall use country names in the address field as an example. Each contribution to *I3* can also be expressed as a percentage.

The observed contributions can be tested against the expected ones on the basis of the distribution of citable items across units of analysis (such as journals or nations). In a larger set, for example, one can expect more highly-cited papers for stochastic reasons. Whether a difference is statistically significant or not can be assessed for each case using the binomial *z*-test or the standardized residuals of the $\chi^2$. We use the latter measure [ $Z = \dfrac{observed - expected}{\sqrt{expected}}$ ] because this test is simpler and less conservative than the binomial test. Expected values below five are discarded as unreliable.

A *z*-value of 1.96 (that is, almost two standard deviations) can be considered as significant at the 5% level, and similarly $z_{0.01}$ = 2.576. The notation of SPSS will be followed in this study using two asterisks for significance at the 1% level, and a single asterisk for the 5% level. However, we use the signs of the differences (++, +, -- or -) when relevant in the tables to indicate whether the observed values are significantly above or below the expected values, and at which level of significance.

In summary, we distinguish between testing (1) observed impacts as sum values of percentiles against expected impacts of units of analysis (e.g., journals, nations, etc.) using *Z*-statistics, and (2) differences in the citation distributions, for example, in terms of Dunn's test. The latter test provides us with a non-parametric alternative to comparing these distributions in terms of their arithmetic averages, as is done in the case of comparing among *IFs* (cf. Leydesdorff, 2008).

**Results**

*I3 for the 65 journals of LIS*

Table 3 provides the Pearson correlations (in the lower triangle) and the Spearman rank-order correlations (upper triangle) between the various indicators under discussion.

**Table 3**: Rank-order correlations (Spearman's $\rho$; upper triangle) and Pearson correlations $r$ (lower triangle) for 65 journals of LIS.

| Indicator | IF-2009 | I3 (100PR) | Mean 100PR | I3 (6PR) | Mean 6PR | Number of publications | Total citations |
|---|---|---|---|---|---|---|---|
| IF-2009 |  | .804 ** | .924 ** | .582 ** | .936 ** | .263 * | .893 ** |
| I3 (100PR) | .591 ** |  | .843 ** | .875 ** | .862 ** | .670 ** | .974 ** |
| Mean 100PR | .839 ** | .651 ** |  | .571 ** | .983 ** | .238 | .907 ** |
| I3 (6PR) | .479 ** | .924 ** | .417 ** |  | .608 ** | .907 ** | .817 ** |
| Mean 6PR | .893 ** | .648 ** | .950** | .506 ** |  | .271 * | .931 ** |
| N of publications | .151 | .619 ** | .042 | .861 ** | .134 |  | .562 ** |
| Total citations | .685 ** | .963 ** | .631 ** | .894 ** | .713 ** | .551 ** |  |

Note: ** Correlation is significant at the 0.01 level (2-tailed); * Correlation is significant at the 0.05 level (2-tailed).

Most of the correlation coefficients are high ($\geq$ .7), and significant at the one percent level. Using Pearson correlation coefficients, however, the numbers of publications (*N*) in journals of this set are not correlated to the *IFs* ($r = .151$; *n.s.*) or the mean values of *100PR* ($r = .042$; *n.s.*) and *6PR* ($r = .134$; *n.s.*). These indicators have in common that they are based on averages and therefore division by the *N* of publications in each set.

The value of *I3*, however, is correlated significantly to both the number of publications ($r = .619$; $p < 0.01$) and the total number of citations ($r = .963$; $p < 0.01$). These correlations are higher than the correlation between the numbers of citations and publications ($r = .555$; ; $p < 0.01$) which is largely a spurious correlation caused by size differences among the 65 journals.

The correlations with both the number of publications and citations could be expected because of the definition of *I3*. Like the *h* index (Hirsch, 2005), *I3* takes both dimensions—the number of publications and their citation rates—into account in the definition of impact, but differently from the *h* index, the tails of the distributions are not discarded as irrelevant. *Ceteris paribus* in terms of the top-segment (e.g, $h = 10$), the number of publications with less impact matters for the overall impact of two otherwise comparable documents sets.

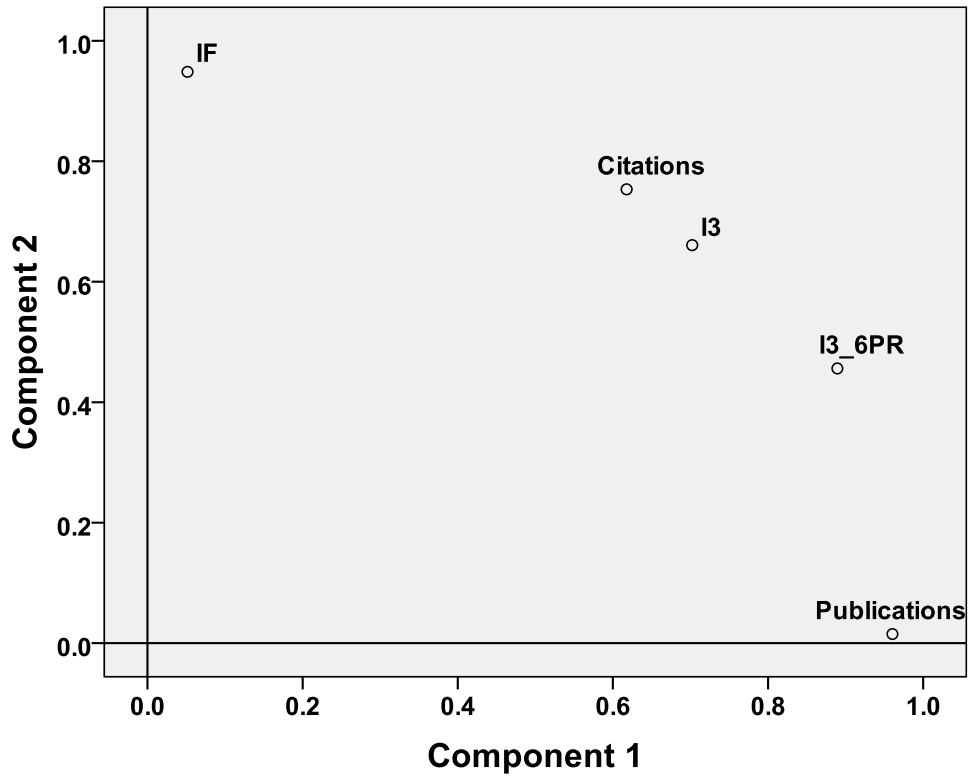

**Figure 5**: Varimax rotated two-factor solution of the variables *IF, I3, I3(6PR)*, number of publications, and citations.

Figure 5 shows the plot of the (Varimax rotated) two-factor solution of the variables that chiefly interest us here. As can be expected, the number of publications and the *IF* span orthogonal coordinates (Leydesdorff, 2009). The *I3* values are closest to total cites because the transformation is linear, whereas a non-linearity is involved in the case of *I3(6PR)*. Unlike the total number of citations, however, the new indicator takes the shapes of the distributions into account by normalizing in terms of percentiles. The number of publications has an effect on *I3* independent of the latter's correlation with the total number of citations.

This can be shown as follows: the partial correlation between *I3* and the number of publications controlled for the number of citations ($r_{I3,N|TC}$) is .391 ($p = 0.01$), whereas $r_{IF,N|TC} = -.373$ ($p < 0.05$). For the *I3(6PR)* this partial correlation $r_{I3(PR6),N|TC} = .985$ ($p < 0.01$) indicating that the binning into six percentile rank classes uncouples relatively from the citation rates and makes the publication rates therefore more important. The hundred percentile ranks provide a finer-grained and therefore more precise indicator of citation impact than the ranking in six classes.

The correlations in Table 3 were calculated at the journal level. Although the correlation between *I3* and the sum of total citations is very high ($r = .963$; $p < 0.01$), the underlying data allow us also to consider the correlation between the times cited and the percentiles at the level of the 5,737 documents. The Pearson correlation is in this case only .639 ($p < 0.01$).[5]

In summary, *I3* provides us with an indicator which takes both the number of publications and their citations into account. The normalization to percentile ranks appreciates the shape of the distribution; the transformation of the citation curve is linear. No parametric assumptions (such as averages and standard deviations) are made. The definitions are sufficiently abstract so that impact is no longer defined in terms of a fixed citation window: any document set can be so evaluated. Different from the $h$ index, the full citation curve is weighted into these non-paramatric statistics.

---

[5] As could be expected, the correlation between times cited and the binned values of *I3(6PR)* is much higher ($r = .815$; $\rho = .911$) because the binning reduces the variance.

**Table 4**: Rankings between 15 journals of LIS with highest values on $I3$ (expressed as percentages of the sum) compared with $IF$s, total citations, and % $I3(6PR)$.

| Journal | N of papers (a) | % I3 (b) | IF 2009 (c) | Total citations (d) | % I3 (6PR) (e) |
|---|---|---|---|---|---|
| J Am Soc Inf Sci Technol | 375 | 9.73 [1]++ | 2.300 [7] | 1975 [1] | 8.63 [1]++ |
| Scientometrics | 258 | 7.24 [2]++ | 2.167 [10] | 1336 [3] | 6.37 [2]++ |
| J Amer Med Inform Assoc | 199 | 6.80 [3]++ | 3.974 [2] | 1784 [2] | 6.15 [3]++ |
| Inform Process Manage | 221 | 6.14 [4]++ | 1.783 [15] | 921 [4] | 4.90 [4]++ |
| Inform Management | 117 | 4.01 [5]++ | 2.282 [8] | 822 [6] | 3.35 [5]++ |
| Int J Geogr Inf Sci | 120 | 3.14 [6]++ | 1.533 [17] | 446 [9] | 2.55 [6]++ |
| MIS Quart | 66 | 2.61 [7]++ | 4.485 [1] | 847 [5] | 2.34 [7]++ |
| J Manage Inform Syst | 82 | 2.60 [8]++ | 2.098 [11] | 496 [8] | 2.31 [8]++ |
| J Health Commun | 90 | 2.52 [9]++ | 1.344 [22] | 380 [10] | 2.04 [10a]++ |
| J Acad Libr | 127 | 2.51 [10]++ | 1.000 [26] | 252 [19] | 2.05 [9] |
| J Inform Sci | 102 | 2.43 [11]++ | 1.706 [16] | 355 [13] | 1.98 [11] |
| J Comput-Mediat Commun | 108 | 2.37 [12]++ | 3.639 [3] | 374 [11] | 2.04 [10b]++ |
| J Informetr | 66 | 2.28 [13]++ | 3.379 [4] | 598 [7] | 2.04 [10c] |
| J Med Libr Assoc | 114 | 2.21 [14]++ | 0.889 [31] | 248 [20] | 1.93 [12] |
| Telecommun Policy | 96 | 2.15 [15]++ | 0.969 [27] | 264 [17] | 1.80 [13] |

Note. ++$p < 0.01$; above the expectation. Ranks are added between brackets.

Table 4 provides the rankings for the 15 journals with the highest values for $I3$ in comparison to rankings of the $IF$s-2009, the total citations, and $I3(6PR)$ as an alternative classification scheme. One can see that on all measures except the IF, *JASIST* is ranked in first place. *MIS Quarterly* holds the seventh position in terms of both $I3$ and $I3(6PR)$.

The highly skewed citation distribution (in column $d$) cannot prevent the *Journal of the American Medical Informatics Association* with 1,784 citations and an *IF* of 3.974 from ranking below *Scientometrics* with only 1,336 citations and the lower *IF* of 2.167, yet nevertheless occupying the second position behind *JASIST*. Below the top segment, the six classes become less fine-grained than the hundred percentiles. This is visible in Table 4 as the *Journal of Health Communication, Journal of Computer-Mediated Communication,* and *Journal of Informetrics* are tied for the tenth position (within this set of 65 journals). In the case of the *Journal of Informetrics*, however, the *I3*(*6PR*) value is no longer significantly different from the expectation.

We have argued that one needs a statistic for testing the differences among citation distributions for their relative significance beyond testing impacts as integrated values against expected impacts. Using the Kruskall-Wallis rank variance test, the null-hypothesis that citation distributions are the same across these 65 journals was rejected at the 1% level. Given this result, we may further test between any two journals whether their citation distributions are significantly different. As noted, we used Dunn's test for the comparison among the citation distributions of the fifty journals with the highest *IF*s (among the 65 in the LIS category); the results are summarized in Figure 6.

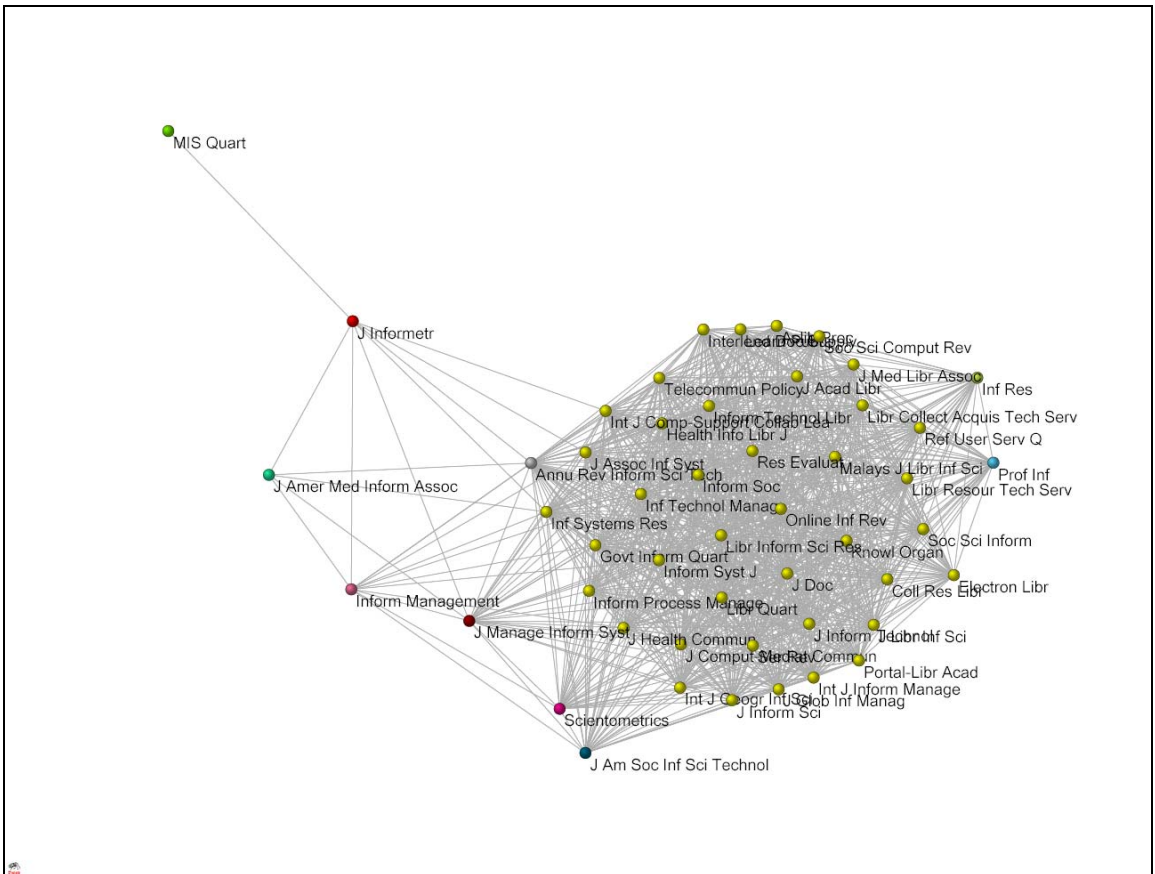

**Figure 6**: Fifty journals of LIS organized according to (dis)similarity in their being-cited patterns to 5,125 publications in 2007 and 2008.
Note. Dunn's test for multiple comparisons ($\alpha < 0.000041 = [0.05 / \{(50 * 49)/2\}]$; Kamada & Kawai (1989) used for the visualization.

Figure 6 shows that *MIS Quarterly* is significantly different in terms of its citation distribution from all other journals in this group except for the *Journal of Informetrics*. (This exceptional distribution leads among other things to the high *IF* of this journal.) Using measures for interdisciplinarity, Leydesdorff & Rafols (2010) have shown that these two journals can be considered as relatively mono-disciplinary specialist journals within this set. *JASIST* and *Scientometrics*—both high on interdisciplinarity relative to this set (!)—are positioned at another corner of the figure (at the bottom) as significantly different from a number of journals in a major group of 37 journals that form a *k* = 25 core set. The *Journal of Computer-Mediated Communication*, for example, is part of this core set with an *IF*-2009 of 3.639, while at the lower end *Interlending & Document Supply* has an *IF*-2009 of 0.403. Differences in *IF*s of an order of magnitude do not inform us about the significance of differences in citation distributions, nor in terms of citation impact unless, of course, one defines "citation impact" in these terms (e.g., Garfield, 1972). Note that *I3* is an indicator of the impact of document sets in terms of citations, and thus the semantics are somewhat different.

*Multidisciplinary Sciences*

The ISI Subject Category MS contains a heterogeneous set of 48 journal, ranging from *Science* and *Nature* with *IF*s of 29.747 and 34.480, respectively, to *R&D Magazine* with *IF* = 0.004 in 2009. However, 65.2% of all citable publications in this set during 2007 and 2008 (that is, 24,494) were published in six major journals: *PNAS* (7,058; 28.8%), *Nature* (2,285; 9.3%), *Science* (2,253; 9.2%), *Annals of the NY Academy of Science* (1,996; 8.2%), *Current Science* (1,271; 5.2%), and the *Chinese Science Bulletin* (1,115; 4.6%).

Among these journals, *Science* and *Nature* seem to have a very similar profile (Figure 7). For example, the number of not-cited papers in this set is 279 for *Science* and 282 for *Nature.* In both cases, this is more than 10% of all citable items in the journal. The largest among these journals, *PNAS*, however, has a very different profile: only 58 (< 1%) of its 7,085 citable publications had never been cited by the date of the download (Feb. 20, 2011).

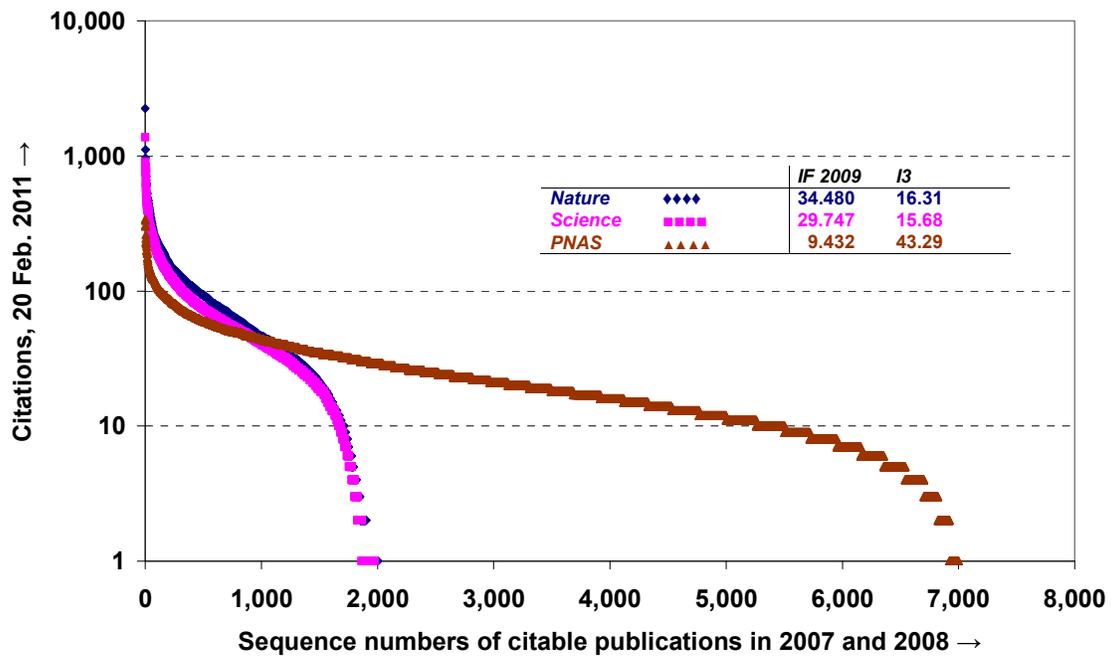

**Figure 7**: Log-scaled citation distributions for the citable publications in 2007 and 2008 in *Nature, Science,* and *PNAS*; downloaded at the WoS on Feb. 20, 2011.

In Figure 7, the numbers of citations are log-scaled in order to make the differences in these skewed distributions more visible. The citation curve for *Nature* remains consistently above the one for *Science*, but the one for *PNAS* is very differently shaped. This journal has in total 27,419 more citations than *Nature*, whereas the latter has 24,488 more citations than *Science* (at this date), yet the *IF* of *PNAS* is less than one-third (*IF-2009* = 9.432). The large tail in the distribution of moderately cited papers works as above to the disadvantage of the larger journal. Note that such a tail would similarly disadvantage a highly productive research team or university.

**Table 5**: Fifteen MS journals with highest values on *I3* compared in ranking with IFs, total citations, and *I3*(6*PR*).

| Journal | N of papers (a) | % I3 (b) | IF 2009 (c) | Total citations (d) | % I3 (6PR) (e) |
|---|---|---|---|---|---|
| *Proc Nat Acad Sci USA* | 7,058 | 43.29 ++ | 9.432 [3] | 178,137 | 33.64 [1] ++ |
| *Nature* | 2,285 | 16.31 ++ | 34.480 [1] | 150,718 | 16.46 [2] ++ |
| *Science* | 2,253 | 15.68 ++ | 29.747 [2] | 126,230 | 15.27 [3] ++ |
| *Ann NY Acad Sci* | 1,996 | 9.33 ++ | 2.670 [5] | 14,284 | 8.29 [4] |
| *Curr Sci* | 1,271 | 2.33 -- | 0.782 [22] | 1,551 | 3.40 [5] -- |
| *Chin Sci Bull* | 1,115 | 2.11 -- | 0.898 [20] | 2,239 | 2.55 [6] -- |
| *Philos Trans R Soc A* | 451 | 1.78 -- | 2.295 [9] | 3,065 | 1.46 [7] -- |
| *J R Soc Interface* | 257 | 1.18 ++ | 4.241 [4] | 2,772 | 0.83 [13] -- |
| *Int J Bifurcation Chaos* | 553 | 1.09 -- | 0.918 [17] | 1,204 | 1.34 [8] -- |
| *Naturwissenschaften* | 288 | 1.01 -- | 2.316 [8] | 1,697 | 0.75 [14] -- |
| *TheScientificWorldJournal* | 358 | 0.82 -- | 1.658 [10] | 1,183 | 0.91 [11] -- |
| *Prog Nat Sci* | 463 | 0.67 -- | 0.704 [24] | 710 | 1.07 [9] -- |
| *Sci Amer* | 370 | 0.37 -- | 2.471 [7] | 585 | 0.87 [12] -- |
| *C R Acad Bulg Sci* | 427 | 0.33 -- | 0.204 [37] | 234 | 0.95 [10] -- |
| *S Afr J Sci* | 191 | 0.33 -- | 0.506 [28] | 220 | 0.47 [15] -- |

Note. ++ $p < 0.01$ above the expectation; -- $p < 0.01$ below the expectation.

Table 5 provides the data for the 15 journals with the highest value of *I3* among the 48 journals of this Subject Category in a format similar to that of Table 4 above (for LIS), and Table 6 provides the correlation coefficients (as above in Table 3). Although the two measures of *I3* and *IF* correlate again significantly over the set (Table 6), the order of the journals as indicated in Table 5 is very different.

For example, *Current Science* and the *Chinese Science Bulletin* were ranked at the 22[nd] and 20[th] place in this set with *IFs* of 0.782 and 0.898, respectively, but are now rated as the fifth and sixth largest impact journals. The scoring in terms of *I3*(*6PR*) in the right-most column of Table 5 shows that this is not only an effect of the large tails in the distribution with infrequently cited papers, but consistent when using this evaluation scheme of the six classes which rewards excellence (top-1%, etc.) disproporionally.

**Table 6**: Rank-order correlations (Spearman's $\rho$; upper triangle) and Pearson correlations $r$ (lower triangle) for the 48 journals of MS.

| Indicator | IF-2009 | I3 | Mean 100PR | I3(6PR) | Mean 6PR | Number of publications | Total citations |
|---|---|---|---|---|---|---|---|
| IF-2009 |  | .798 ** | .884 ** | .517 ** | .844 ** | .479 ** | .840 ** |
| I3 | .590 ** |  | .777 ** | .854 ** | .837 ** | .829 ** | .986 ** |
| Mean 100PR | .775 ** | .691 ** |  | .408 ** | .817 ** | .364 * | .838 ** |
| I3(6PR) | .660 ** | .987 ** | .706 ** |  | .646 ** | .996 ** | .801 ** |
| Mean 6PR. | .956 ** | .716 ** | .875 ** | .775 ** |  | .605 ** | .887 ** |
| N of publications | .492 ** | .953 ** | .605 ** | .967 ** | .635 ** |  | .772 ** |
| Total citations | .841 ** | .922 ** | .756 ** | .945 ** | .884 ** | .839 ** |  |

Note: ** Correlation is significant at the 0.01 level (2-tailed); * Correlation is significant at the 0.05 level (2-tailed).

Table 6 shows that in this case, the differences in size among these 48 journals are so important that all correlations are much higher. In other words, these correlations are spurious since they are caused by differences in size more than in the previous case since there are six major journals and 42 small ones. Despite the higher correlations, the same effects as discussed above for the case of LIS can be found. The partial correlation between *I3* and the number of publication controlled for the number of citation $r_{I3, N|TC}$ = .850 ($p < 0.01$), whereas $r_{I3(6PR), N|TC}$ = .982 ($p < 0.01$) is again higher. As in the previous case, $r_{IF, N|TC}$ is negative (-.724; $p < 0.01$) because the *IF* is based on dividing by the *N* of publications.

Both *I3* and *I3*(*6PR*) correlate again significantly with the number of citable publications and citations. This follows from the definition of impact by analogy to the product of "mass" (number of publications) and "velocity" (quality of each publication). Both terms thus contribute to the impact, but with qualification of the citedness of each publication in terms of its percentile rank.

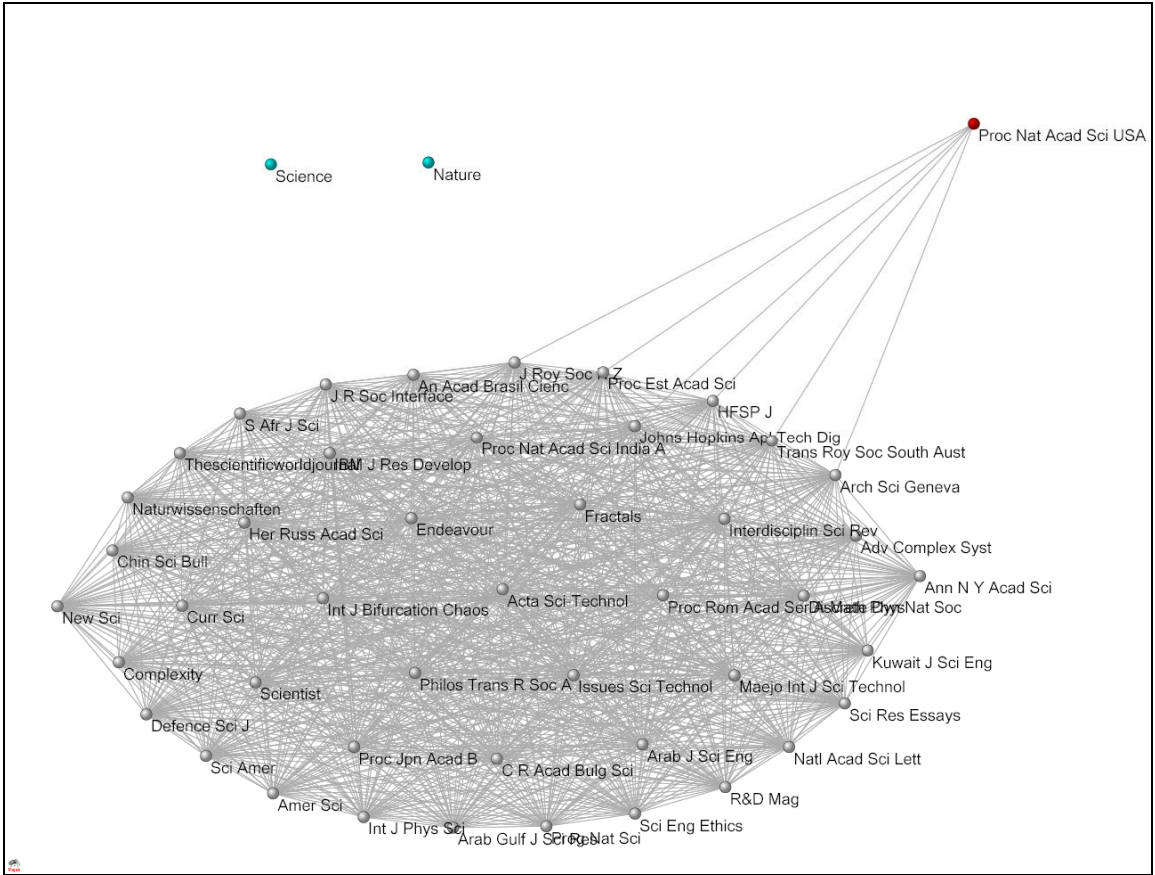

**Figure 8**: Forty-eight journals in MS organized according to (dis)similarity in their being-cited patterns to 24,494 publications in 2007 and 2008.
Note. Dunn's test for multiple comparisons ($\alpha < 0.000044 = [0.05 / \{(48 * 47)/2\}]$; Kamada & Kawai (1989) used for the visualization.

Dunn's test applied to the citation patterns of these 48 journals are visualized in Figure 8. The figure shows that 45 of the 48 journals form a *k*=25 core set of journals. Both *Science* and *Nature* are significantly different in their citation patterns from all other journals in this set, and—perhaps counter-intuitively—from each other. The citation distribution of *PNAS*, however, is not significantly different from a number of other journals in the set.

*Current Science* (Curr Sci) and the *Chinese Science Bulletin* (Chin Sci Bull), however, are positioned to the left within the core-set, in the neighbourhood of the *New Scientist* (New Sci). This latter journal has an *IF* of only 0.333 and a contribution to the impact in terms of *I3* of only 0.17% of the total impact of the set (1,035,332.14).

In summary, these results show that, on the one hand, two journals with similarly high *IFs* such as *Science* and *Nature,* can nevertheless differ significantly in their citation distributions. Note that Dunn's test is performed directly on the raw citation scores, that is, before the transformation into percentile ranks. On the other hand, a journal with a very high impact such as *PNAS* may not differ much in its citation pattern from a journal like the *Proceedings of the Estonian Academy of Science,* although both their respective impacts *I3* and impact factors *IFs* differ by orders of magnitude.

Let us recall that the *IF* was designed precisely with the objective to correct for these size differences between otherwise similar journals such as *PNAS* and the *Proceedings of the Estonian Academy of Science* (Bensman, 2008; Garfield, 1972). It completely fails to do so because of the parametric assumption involved in using an arithmetic average (cf. Rousseau & Leydesdorff, 2011).

**Performance measurement**

Because percentiles are attributed at the paper level, the datasets enable us to perform aggregations and decompositions other than in terms of journals or journal sets. Documents and document sets can both be analyzed in terms of disciplinary structures and be considered as products of authors, institutions, nations, etc. (Narin, 1976; Small &

Garfield, 1985). On the one side, one refers to what a journal accepts as worthy of publication, whereas, the other refers to how a person, say, performs and communicates scientific work. Whereas journals do not produce scientific knowledge the way people/instiitions do and, of course, are evaluated, *I3* is so general that one is enabled to combine the two different types of evaluations (Leydesdorff, 2008). One can, for example, compare the productivity and impact of an institution or nation in two different journal categories (and test the difference for its significance.)

Let us as an example recompose the 5,737 citable documents of the LIS set using the country addresses provided in the bylines of these papers. Fractional counting of the addresses will be used in order to keep the total numbers consistent. In other words, if a paper is coauthored between two authors from country A and one from country B, the attribution is for one-third to country B and two-thirds to country A.

Table 7 provides the results; the table is composed by first using only players in the field with at least a one-percent contribution to *I3*, and then sorted in column c using the ratio of this share divided by the percentage share of publications as the expected distribution (column a). (The regression line is in this case less interesting since overdetermined by the outliers for the USA and EU 27.)

**Table 7**: Percentages shares of publications (2007 and 2008) and %*I3* in the set of 65 journals of LIS (sorted by the ratio of percentage of *I3* / percentage share of publications in column c).

|  | Percentage publications (a) | % *I3* (b) | Ratio of (b) and (a) (c) | % *I3*(*6PR*) (d) | Ratio of (d) and (a) (f) |
|---|---|---|---|---|---|
| Netherlands | 2.23 | 3.75 | 1.68 ++ | 3.23 | 1.45 ++ |
| Switzerland | 0.77 | 1.24 | 1.61 ++ | 1.21 | 1.57 ++ |
| Belgium | 1.15 | 1.81 | 1.57 ++ | 1.56 | 1.36 ++ |
| South Korea | 1.45 | 1.88 | 1.30 ++ | 1.58 | 1.09 + |
| Taiwan | 2.23 | 2.87 | 1.29 ++ | 2.52 | 1.13 ++ |
| Peoples R China | 2.35 | 2.97 | 1.26 ++ | 2.63 | 1.12 ++ |
| Italy | 0.83 | 1.02 | 1.23 ++ | 0.86 | 1.04 |
| EU-27 | 24.67 | 30.21 | 1.22 ++ | 28.20 | 1.14 ++ |
| Canada | 3.46 | 4.08 | 1.18 ++ | 3.73 | 1.08 ++ |
| Australia | 2.19 | 2.58 | 1.18 ++ | 2.37 | 1.08 + |
| Singapore | 1.16 | 1.34 | 1.16 ++ | 1.17 | 1.01 |
| UK | 8.82 | 10.18 | 1.15 ++ | 9.49 | 1.08 ++ |
| Sweden | 0.96 | 1.06 | 1.10 | 1.04 | 1.08 |
| USA | 40.36 | 42.91 | 1.06 ++ | 41.52 | 1.03 ++ |
| France | 1.10 | 1.16 | 1.05 ++ | 1.13 | 1.03 |
| Finland | 1.09 | 1.10 | 1.01 | 1.05 | 0.96 |
| Spain | 4.31 | 4.22 | 0.98 - | 4.07 | 0.94 |
| Germany | 2.51 | 2.38 | 0.95 -- | 2.58 | 1.03 |
| % accounted | 88.72 | 97.58 | 1.10 ++ | 92.91 | 1.05 ++ |

Note. ++ $p < 0.01$ above the expectation; -- $p < 0.01$ below the expectation; + $p < 0.05$ above the expectation; - $p < 0.05$ below the expectation.

Only 5,090 (88.72%) of the 5,737 records contained (8,510) addresses with country names. These records are cited more often than records without addresses, so that the "world average" given the set of 65 LIS journals would be 1.10 using *I3,* or 1.05 in the case of *I3(6PR)*. Because this indicator is based on a summation, the value for the EU-27 is equal to the sum value of the 27 nations composing the EU. (Similarly, the value for the UK is constructed by adding records with England, Scotland, Wales, and Northern Ireland as country indicators in the ISI set.)

Using *I3*, the Netherlands scores highest with 1.68, but using *I3*(*6PR*) Switzerland which was second on the scale of one hundred, is now highest with 1.57. As noted, *I3*(*6PR*) is sensitive to an above-average representation in the top segments of the percentile distribution. Switzerland is known to be well represented in these segments (e.g., King, 2004). The USA outperforms all other nations (including a constructed EU-27) in terms of absolute numbers on both scales. The low contribution of the PR China in this set is notable. Important countries such as India, Russia, and Brazil are not listed as contributing because of the threshold used of more than a 1% contribution to the total impact.

In summary, percentile ranks are defined at the level of each individual paper in a set. How the set is composed (for example, in terms of publications in two years in this study, but for the purpose of a comparison with the *IFs*) can still be decided on the basis of a research question. For example, one may wish to compare the impact of publications of rejected versus granted PIs in a competition (Bornmann *et al*., 2010; Van den Besselaar & Leydesdorff, 2009). In each such study, one can determine percentiles, test citation curves against one another for the statistical signifance of differences (using Dunn's test), and test for each subset whether the impact is significantly above of below the expectations (using the *Z*-test). Our method is thus most general and avoids parametric assumptions.

**Conclusions and discussion**

We elaborated above on the *I3* values using the two-year set of citable items in order to facilitate—in this study—the comparison with the *IF*. However, as shown above, this indicator is not restricted to journals, document sets, time-periods, etc., but more general:

only the specification of a reference set is required from which the samples under study are drawn (Bornmann *et al.*, 2008). Above, we used two ISI Subject Categories as reference sets, but one could also use the entire *Science Citation Index*, *Scopus* data, data from Google Scholar, or patent databases that contain citations. One can even apply this to a citation count in "grey literature." *I3* provides a general measure of citaiton that can be applied across samples of different sizes; the non-parametric statistics account for the typically highly-skewed citation distributions.

Our first point was that impact is not captured correctly using a central tendency statistics such as the mean or the median. Lundberg (2007, at p. 148) noted that one does not have to average the (field-normalized) citation scores, but can also use their sum values as a "total" field-normalized citation score. Using the Leiden Rankings, the Center for Science and Technology Studies (CWTS) multiplied the product of the number of publications *P* with the "old" crown indicator *CPP/FCSm* in order to obtain as a result what was called the "brute force indicator." In the new set of Leiden indicators, analogously, a "total normalized citation score" is proposed (Van Raan *et al.*, 2010b, at p. 291). However, all these indicators are based on the parametric assumption (of the Central Limit Theorem) that one is allowed to compute with the mean as a summary statistics given a sufficiently large number of observations (e.g., Glänzel, 2010).

As with the impact factors, citation analysis has hitherto been caught in the paradigm of parametric statistics, although this approach is mostly not fruitful for bibliometrics (cf. Ahlgren *et al.*, 2003). Changing to the median, however, is not sufficient because the median as a central-tendency measure is as sensitive—and sometimes even more so—to the tails of the distributions. A finer-grained scheme of hundred percentiles can be envisaged. Actually, we used the percentile ranks above as a continuous random variable which can be specified to any desirable degree of precision in terms of decimal numbers. Thus defined, the percentile ranks are attributes of the publications which can be added in order to perform integration along the qualified citation curve.

Additionally, we showed that one can vary the evaluation scheme using the six percentile ranks that are used in the *Science and Engineering Indicators* (National Science Board, 2010, Appendix Table 5-43; Bornmann & Mutz, 2011). The emphasis on the more-highly-cited publications in this scheme enhances the distinctions as more significant (e.g., in Tables 5 and 7), but one may lose some information such as fine-grained distinctions between units of analysis with tied ranks. *I3* for hundred percentiles provides the general scheme from which others can be derived given different policy contexts. As noted, hundred percentiles can be considered as a continuous variable, and one can thus provide the degree of precision in decimals.

In the meantime, the percentile rank approach is also used by the new InCites database of Thomson Reuters that functions as an overlay to the Web of Science. Unfortunately, the percentile ranks are averaged in this case and one cannot escape from the scheme of ISI Subject Categories as the reference sets for determining the percentiles (cf. Pudovkin & Garfield, 2002). Using percentile ranks, however, the classification into categories can in the future also be paper-based, such as using the Medical Subject Heading (MeSH) in the Medline database of the NIH (Bornmann *et al*., 2008) or using the keywords of dedicated databases such as Chemical Abstracts (Bornmann *et al*., 2011). We expect the state of the art to change rapidly in this respect.

Our suggestion to use summations for the impact may raise the question of whether impact per paper was not defined above as rate of summations rather than a summation of rates. Last year's debate about normalization was about using "rates of averages" versus "averages of rates," as Gingras & Larivière (2011) succinctly summarized the crucial issue of the controversy. However, percentile ranks are rates, albeit non-parametric ones. As shown above (e.g., in Figure 4), the resulting sums for different units of analysis can be regressed upon the number of publications and thus the impact/paper can be indicated. This impact/paper can be tested for its significance against the distribution of papers under study (using $\chi^2$-square statistics). The differences in underlying citation distributions can be tested for their significance using, for example, Dunn's test. Using percentiles, the evaluation scheme for both the performance of authors and institutes, and the quality of journals can methodologically be brought into a single framework. Since the *I3* measure in fully decomposable, multi-dimensional distinctions are also possible.

In terms of the statistics, our main message is to keep significance in differences among citation distributions analytically separate from impact, which we defined—in analogy to the (vector-)summation of momenta in physics—as summations of products. Thus defined, the percentile rank approach of the Integrated Impact Indicator (*I3*) enables us to take both the size and the shape of the distribution into account, and impacts among different units of analysis (journals, nations, universities, institutions, individuals) can be compared (that is, added and substracted) as percentages. Whether these units of analysis (e.g., individuals, research groups, small countries such as Monaco) are large enough for the comparison can be decided on statistical (instead of normative or moral) grounds. Differences which are not significant can be dis-advised for usage in policy making and research management.

**Policy implications**

The common assumption in citation impact analysis hitherto has been normalization to the mean. In our opinion, the results are then necessarily flawed because the citation distributions are often highly-skewed. Highly productive units can then be disadvantaged because they publish often in addition to higher-cited papers also a number of less-cited ones which depress their average performance. We became aware of this when we tried to reproduce the performance of seven PIs of the Academic Medical Center of the University of Amsterdam whom had been evaluated by CWTS. The first and sixth position among these seven were swapped when we used mean percentile ranks (Leydesdorff et al., 2011). Thus, the effect of the proposed change in the paradigm of impact assessment can be highly significant, both in terms of the statistics and policy implications. In this case, for example, the ranking in terms of impact was used as input to the funding scheme of these PIs; the research group of the sixth PI thus suffered a loss in funding because of this group's productivity other than in the top-1% (*personal communication*, November 2010).

In this study, we generalized the non-paramatric approach. Long-standing assumptions such as *Nature* and *Science* would have higher citation impact (given higher *IF*s) when compared with *PNAS*, were shown above as erroneous consequences of parametric assumptions. *Current Science* and the *Chinese Science Bulletin* were ranked at the 22$^{nd}$ and 20$^{th}$ place in the set of multidisciplinary journals with *IF*s of 0.782 and 0.898, respectively, but were rated as the fifth and sixth largest impact journals using *I3*. Thus, this analysis in terms of percentiles shows the increased importance of these two multidisiciplinary journals, while the *IFs* do not.

Narin's (1976) original scheme of crosstabling journals and nations in evaluative bibliometrics can be considered as two dimensions for the aggregation into subsets of papers. *I3* is additive and therefore allows for the comparison of subsets which may differ along both axes. Actually, one can also compute a contribution of *I3* for a set that differs in terms of both addresses and journals (or other parameters) given that the reference sets for the percentile ranks are properly set as the relevant total sum sets. Whereas the *h* index and its derived statistics seem to allow for such comparisons, *I3* does not simplify the computation by discarding the bulk of the references in the tails of the distributions. Note that papers with zero citations do not contributed to *I3*, because *I3* is an impact indicator that takes both publication quantities and (normalized) citation impact into account. Using *I3* for the evaluation, one is no longer "punished" for one's productivity.

**Acknowledgements**

We are grateful to Rüdiger Mutz for comments on previous drafts.